\documentclass{article}
\usepackage{spconf,amsmath,graphicx}

\usepackage{algorithm}
\usepackage[noend]{algpseudocode}
\usepackage{hyperref}
\usepackage{flexisym}
\usepackage[activate={true,nocompatibility},spacing=true,final,tracking=true,kerning=true,factor=1100,stretch=10,shrink=10]{microtype}

\title{Fine-tuning of Pre-trained End-to-end Speech Recognition with Generative Adversarial Networks}
\name{Md. Akmal Haidar and Mehdi Rezagholizadeh}
\address{Huawei Noah's Ark Lab, Montreal Research Centre, Canada \\
{\fontsize{10pt}{10pt}\selectfont \{md.akmal.haidar, mehdi.rezagholizadeh\}@huawei.com}}

\begin{document}
\maketitle

\ninept
\begin{abstract}

Adversarial training of end-to-end (E2E) ASR systems using generative adversarial networks (GAN) has recently been explored  for low-resource ASR corpora. GANs help to learn the true data representation through a two-player min-max game. However, training an E2E ASR model using a large ASR corpus with a GAN framework has never been explored, because it might take excessively long time due to high-variance gradient updates and face convergence issues. In this paper, we introduce a novel framework for fine-tuning a pre-trained ASR model using the GAN objective where the ASR model acts as a generator and a discriminator tries to distinguish the ASR output from the real data. Since the ASR model is pre-trained, we hypothesize that the ASR model output (soft distribution vectors) helps to get higher scores from the discriminator and makes the task of the discriminator harder within our GAN framework, which in turn improves the performance of the ASR model in the fine-tuning stage. Here, the pre-trained ASR model is fine-tuned adversarially against the discriminator using an additional adversarial loss. Experiments on full LibriSpeech dataset show that our proposed  approach outperforms baselines and conventional GAN-based  adversarial models.

\end{abstract}
\vspace{-4mm}
\section{Introduction}
\label{sec:intro}
End-to-end (E2E) automatic speech recognition (ASR) systems map speech acoustic signal to text transcription by using a single sequence-to-sequence neural model without decomposing the problem into different parts such as lexicon modeling, acoustic modeling and language modeling as in traditional ASR architectures~\cite{povey}. It has received a lot of attention because of its simple training and inference procedures over traditional HMM-based systems which require a hand-crafted pronunciation dictionary and a complex decoding system using a finite state transducer (FST)~\cite{fst}. One of the earliest E2E ASR model is the connectionist temporal classification (CTC)~\cite{graves2006connectionist} model which independently maps acoustic frames into outputs. To get better results with CTC, the CTC output needs to be rescored with language models~\cite{graves2014towards}. The conditional independence assumption in CTC was tackled by recurrent neural network transducer (RNNT) model~\cite{graves2012sequence, He2018} which showed better performance for streaming. Attention based encoder-decoder networks yield state-of-the-art results for offline ASR model~\cite{chan2016listen,karita2019is, karita2019asru}. These networks are trained by using sequence-to-sequence and/or CTC losses to learn the true data distribution. 

Generative adversarial networks (GAN)~\cite{goodfellow2014generative} provides another way to learn the true data distribution in a minimax game using two networks: generator and discriminator. A generator generates labels while the discriminator tries to distinguish the true labels from the generated ones. The generator learns from the discriminator via adversarial loss to model the true data distribution. However, the adversarial training using GAN requires a large number of training samples and epochs to converge~\cite{feizi}. This might be because the gradients from the discriminator to update the generator often vanish or explode during the adversarial training~\cite{gan}. 

GANs have been recently explored for robust ASR which showed that inducing invariance at the encoder embedding level improves the recognition of simulated far-field speech recognition~\cite{robust}. GANs have been investigated extensively for speech enhancement~\cite{segan, segan1}, speech dereverberation~\cite{dereverberation}.
In~\cite{aipnet}, an accent-invariant pre-training network was trained using GAN for E2E ASR. In~\cite{nonvc},  a non-parallel voice conversion approach with CycleGAN for speaker adaptation was proposed to improve the ASR performance. 
Adversarial training for E2E ASR using GAN has recently studied  using a low-resource paired ASR corpus with un-paired speech and text corpora~\cite{liu,semi}. In~\cite{semi}, non-parallel speech and text corpora was used to learn a semi-supervised ASR model using adversarial training. In~\cite{liu}, adversarial training was employed for a small paired speech corpus (LibriSpeech 100 hours) and also incorporated unpaired text data to better utilize additional text data and avoid the use of a separately trained language model. However, adversarial training using GANs has been never explored for large paired speech corpora. 

In this work, we investigate adversarial training of ASR using GANs for a large paired speech corpus (LibriSpeech 960 hours). Since adversarial training requires large number of training epochs due to high variance gradient updates~\cite{feizi,gan}, in this work, we utilize the GAN objective for fine-tuning an E2E ASR model, which is pre-trained with a large paired speech corpus. First, the ASR network is trained to learn the true data distribution using cross-entropy and CTC losses~\cite{karita2019asru}. After pre-training, the ASR model would give a smoother soft output representation which can help in training a stronger discriminator to improve the performance of the ASR model (generator). 
We perform extensive experiments using the full librispeech corpus and show that our fine-tuning approach using GAN outperforms baselines and conventional adversarial training of E2E ASR without pre-training.  


\vspace{-3mm}
\section{Proposed Approach}
\subsection{Adversarial Training}
In the adversarial training of end-to-end ASR model, the ASR model acts as a generator conditioned on the speech signal and predicts the corresponding transcription. The discriminator tries to distinguish the real (ground-truth) transcriptions from the ASR output transcriptions. It learns to give higher scores to real texts and lower score to the ASR transcriptions during training. In this paper, we apply   adversarial training to fine-tuining a pre-trained E2E ASR model, which is trained using sequence-to-sequence (s2s) and CTC losses. During fine-tuning the ASR model, the discriminator parameters are fixed and the ASR model is trained by an additional adversarial loss and generate ASR transcriptions to fool the discriminator (i.e., they can get higher score from the discriminator). The ASR model and the discriminator are trained alternately and learn from each other step by step. The algorithm for the adversarial fine-tuning of a pre-trained ASR model is described in Algorithm~\ref{alg:AE-GAN}. In the following subsections, we describe the discriminator and the ASR network architectures. Then, we describe the fine-tuning of the ASR model using GAN.    
\vspace{-2mm}
\subsection{Discriminator Network}
\label{sec:critics}
\begin{figure}[!htb]
  \centering
  \includegraphics[scale=0.9]{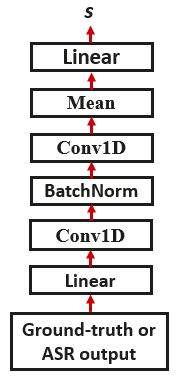}
  \caption{Discriminator Network}
  \label{lmcritic}
  \vspace{-3mm}
\end{figure}

The input to the discriminator (D) network is either the ground-truth (one-hot vectors) or the ASR output transcriptions (soft distribution vectors) and it returns a scalar $s$ as a  quality score. An example discriminator network that we use in our experiments is depicted in Fig.~\ref{lmcritic}. The real text $Y$ or the ASR output $\hat{Y}$ is first mapped to a lower dimension 128 by using a  linear transformation. Then, two one-dimensional convolutional neural network (Conv1D) layers with $128$ kernels, kernel size $2\times1$, and stride $1$ are applied to extract the features for each time index. Batch normalization is applied between layers. 
Finally, the mean feature is calculated over the time axis which is then mapped to a single scalar valuse $s$ with linear projections~\cite{liu}. 

To train the discriminator, we incorporate an improved version  of Wasserstein GAN (WGAN) approach~\cite{wgan} with gradient penalty (gp)~\cite{gulrajani2017improved}. Here, the  discriminator is designed to estimate the Earth-Mover (Wasserstein-1)~\cite{earth} distance between the ground-truth transcriptions and the ASR output transcriptions. 
The loss function of the discriminator can be defined as~\cite{liu}: 
\begin{equation}
L_{D}= \lambda_d(\underset{\hat{Y} \sim P_\theta}{E} [D(\hat{Y})] - \underset{Y \sim P_r}{E} [D(Y)]) + \lambda_{gp} gp   
\end{equation}
where, $\lambda_d$ and $\lambda_{gp}$ are weights. $D(\hat{Y})$ is the scalar output of the discriminator for ASR output $\hat{Y}$. $P_\theta$ and $P_r$ are the distributions for the ASR output and ground-truth (real) transcriptions respectively. For the gradient penalty $gp$ term, we need to calculate the gradient norm of random samples $\Bar{Y}\sim P_{\Bar{Y}}$. gp is calculated as following~\cite{gulrajani2017improved}:
\begin{equation}
gp= \underset{\Bar{Y} \sim P_{\Bar{Y}}}{E} [(||\nabla_{\Bar{Y}} D(\Bar{Y})||_2-1)^2]\\  
\end{equation}
where $\Bar{Y}$ are random samples which can be obtained by sampling uniformly along the line connecting pairs of $\hat{Y}$ and $Y$ samples:  
\begin{equation}
[\Bar{Y}\sim P_{\Bar{Y}}] \leftarrow \gamma ~ [Y\sim P_{r}] + (1-\gamma)~ [\hat{Y} \sim P_\theta]
\end{equation}
where $\gamma \sim U[0,1]$~\cite{gulrajani2017improved}.

\subsection{ASR Architecture}
\begin{figure}[!htp]
  \centering
  \includegraphics[scale=0.95]{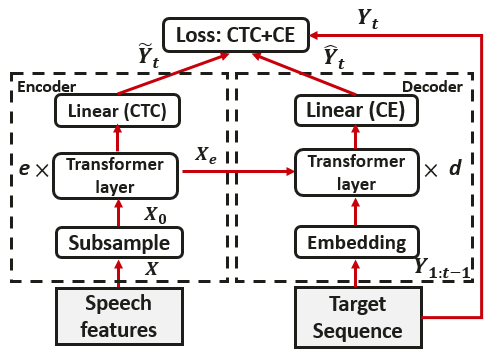}
  \caption{ASR Model Architecture}
  \label{asr}
\end{figure}

Our proposed approach can be applied to any type of ASR network architecture. In this paper, we use a transformer-based ASR network with joint CTC and attention~\cite{karita2019asru, karita2019is}. A schematic of the ASR model is shown in Fig.~\ref{asr}. Given an input sequence $X$ of log-mel filterbank speech features, Transformer predicts a target sequence $\hat{Y}$ of characters or SentencePiece~\cite{kudo2018sentencepiece}. In Fig.~\ref{asr}, the subsample block transforms the source sequence $X$ into a subsampled sequence $X_0 \in R^{n_{sub}\times d_{att}}$ by using a two-layer convolution neural network (CNN) block~\cite{karita2019asru,moritz2020}. Here, $n_{sub}$ is the length of the subsampled sequence and $d_{att}$ is the dimensions of the features~\cite{karita2019asru}. Both CNN layers use a stride of size 2, a kernel size of 3 × 3, and a ReLU activation function. Thus, the striding reduces the frame rate of output sequence $X_0$ by a factor of 4 compared to the feature frame rate of $X$~\cite{moritz2020}. Then, it is followed by a stack of $e$ transformer layers that transform $X_0$ into a sequence of encoded features $X_e \in R^{n_{sub}\times d_{att}}$ for the CTC and decoder networks~\cite{karita2019asru}. The encoder transformer  layers iteratively refine the representation of the input sequence with a combination of multi-head ($h$) self-attention (MHA) and position-wise feed forward networks (FFN) with dimensions $d_{ff}$. 

The decoder generates a transcription sequence $Y=(Y_1,...,Y_{t})$ one token at a time. Each choice of output token $Y_t$ is conditioned on the encoder representations $X_e$ and previously generated tokens $(Y_1,...,Y_{t-1})$ through attention mechanisms. Each decoder layer performs two rounds of multi-head attention: the first one being self-attention over the representations of previously emitted tokens, and the second being attention over the output of the final layer of the encoder. The multi-head attention is then followed by the position-wise FFN~\cite{karita2019asru}.  For the decoder input $Y_{1:t-1}$, we use ground-truth labels in the training stage, while we use generated outputs in the decoding stage. The output of the final decoder layer for the token $Y_{t-1}$ is used to predict the following token $Y_t$. The details of the MHA, FFN, and other components of the architecture such as sinusoidal positional encodings, residual connections and layer normalization are described in~\cite{vaswani2017attention,karita2019asru}. The positional encodings are applied into $X_0$ and $Y_0$ when convolutional 2d subsampling was used~\cite{karita2019asru,moritz2020}.

During ASR training, the frame-wise posterior distribution of $P_{s2s}(\hat{Y}|X)$ and $P_{ctc}(\Tilde{Y}|X)$ are predicted by the decoder and the CTC module respectively. Bear in mind that the discriminator takes only $\hat{Y}$ as input. The ASR model is trained by minimizing the loss function~\cite{karita2019asru}:

\begin{equation}
L_{ASR}=L_{s2s}+L_{ctc}
\end{equation}
where $L_{s2s}=-(1-\alpha)\log P_{s2s}(\hat{Y}|X)$, $L_{ctc}=-\alpha \log P_{ctc}(\Tilde{Y}|X)$, and $\alpha$ is a hyperparameter.  
In the decoding stage, given the speech feature $X$ and the previous predicted token, the next token is predicted using beam search, which combines the scores of the sequence-to-sequence (s2s) and CTC with/ without language model score following~\cite{karita2019asru}. 

\vspace{-1mm}
\subsection{Fine-tuning ASR Model Using GAN}
\begin{figure}[!htb]
  \centering
  \includegraphics[scale=0.9]{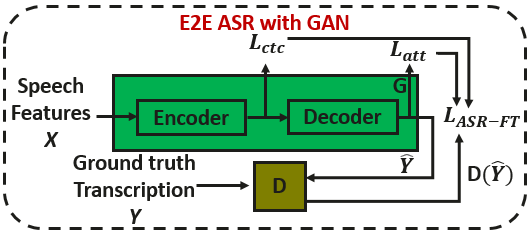}
  \caption{Fine-tuning E2E ASR with GAN}
  \label{ftgan}
\end{figure}

The schematic of the adversarial fine-tuning using GAN is  shown in Fig.~\ref{ftgan}. In this figure, the encoder-decoder architecture is a pre-trained ASR model which acts as a generator ($G$). The discriminator $D$ tries to distinguish the ASR transcriptions from ground-truth transcriptions. Since the ASR model is pre-trained, the discriminator cannot easily discriminate these two transcriptions, which will help to train a stronger discriminator. The discriminator and the generator (ASR model) are trained alternately. The discriminator is trained by minimizing the loss function $L_D$ described in section~\ref{sec:critics}.   
The loss function for fine-tuning the ASR model can be described following~\cite{liu}:
\begin{equation}
L_{ASR-FT}=L_{s2s}+L_{ctc} - \lambda_d D(\hat{Y})    
\end{equation}
where $L_{s2s}$ and $L_{ctc}$ are the sequence-to-sequence and the CTC losses respectively. The term $-\lambda_d  D(\hat{Y})$ represents the adversarial loss which helps the ASR model to get the high quality score from the discriminator. 
The fine-tuning procedure using adversarial training is explained in Algorithm~\ref{alg:AE-GAN}.   

\begin{algorithm}
\caption{Adversarial training for fine-tuning a pretrained ASR Model} 

\label{alg:AE-GAN}
\begin{algorithmic}[1]
\Require{The Adam hyperparameters $\alpha_l$, $\beta_1$, $\beta_2$, $\epsilon, $batch size $m$, pre-trained ASR model parameters $\theta$, and initial discriminator parameters $w_0$}

\For {\text{number of training epochs}}
\For {\text{number of training iterations}}
\State {Sample labels $\{Y^{(i)}\}_{i=1}^m \sim P_r$ and  speech input  \Statex \hspace{1cm} features $\{X^{(i)}\}_{i=1}^m \sim P_X$. \\ \hspace{1cm}predict ASR transcriptions $\{\hat{Y}^{(i)} \}_{i=1}^m \sim P_\theta$. 
\Statex \hspace{1cm}\textbf{Train the discriminator:}
\\ \hspace{1cm} Backpropagate discriminator loss $L_{D}(w)$ . \\
\hspace{1cm}Update with $w\gets Adam(L_{D}(w), \alpha_l, \beta_1, \beta_2, \epsilon) $.
\Statex	\hspace{1cm}\textbf{Fine-tune the ASR Model:} 
\\ \hspace{1cm} Backpropagate ASR loss $L_{AST-FT}(\theta)$. \\
\hspace{1cm} Update with $\theta\gets Adam(L_{ASR-FT}(\theta), \alpha_l, \beta_1, \beta_2,\epsilon) $}.
\EndFor $\textbf{end for}$
\EndFor $\textbf{end for}$
\end{algorithmic}
\end{algorithm}

\vspace{-5mm}
\section{Experiments}
\subsection{Data and Setup}
\label{sec:experiments}
We use the open-source, ESPNet toolkit~\cite{espnet} for our experiments. We conduct our experiments on LibriSpeech dataset~\cite{panayotov2015librispeech}, which is a speech corpus of reading English audio books. It has 960 hours of training data, 10.7 hours of development data, and 10.5 hours of test data, whereby the development and the test data sets are both split into approximately two halves named “clean” and “other”. To extract the input features for speech, we follow the same setup as in~\cite{espnet, karita2019asru}: using 83-dimensional log-Mel filterbanks frames with pitch features~\cite{karita2019asru,moritz2020}. The output tokens come from a 5K sub-word vocabulary created with sentencepiece~\cite{kudo2018sentencepiece}``unigram''~\cite{espnet}. We perform experiments on two settings: small ($e=12, d=6, d_{ff}=2048, d_{att}=256, h=4$) with around 30 M model parameters and large ($e=12, d=6, d_{ff}=2048, d_{att}=512, h=8$) with around 75 M model parameters. We apply default settings of SpecAugmentation~\cite{specaugment} of ESPnet~\cite{espnet}. For small settings, we use batch size of 140, and gradient accumulation of 2. For large settings, we use batch size of 100 and gradient accumulation of 4~\cite{espnet}. All the experiments were run on 4 Tesla V100 gpus. We use Adam optimizer with learning rate scheduling similar to~\cite{vaswani2017attention,espnet} and other  settings (e.g., dropout, warmup steps, $\alpha$, learning rate, label smoothing penalty~\cite{specaugment}) following~\cite{moritz2020,espnet}. 
We train the baseline experiments with small and large settings for 100 and 120 epochs respectively~\cite{espnet}. We average the best five checkpoints as the final model and report results based on it. For decoding, the CTC weight, the LM weight and beam size are 0.5, 0.7 and 20 respectively for the small settings, and 0.4, 0.6, and 30 for the large setup~\cite{moritz2020}.

For adversarial fine-tuning with GAN experiments, we use our trained best average model for initialization and then train within GAN framework for 50 epochs with the Adam optimizer with fixed parameters of $\alpha_l=0.0001$, $\beta_1=0.5$, $\beta_2=.98$, $\epsilon=10^{-9}$~\cite{espnet}. We set $\lambda_{gp}=10$, $\alpha=0.3$, and 
$\lambda_d=0.0001$ as the discriminator output value $s$ was usually higher than other loss values~\cite{liu}.
For our proposed fine-tuning experiments, we see that the CTC weight of 0.4 and insertion penalty of 2 give better results with LM fusion over the baseline models for both small and large settings.  
For LM fusion~\cite{shallowfusion}, we use a pretrained  Transformer LM for decoding\footnote{ \url{https://github.com/espnet/espnet/blob/master/egs/librispeech/asr1/RESULTS.md}}.  

\begin{table}[!htb]
  \small
  \centering
  \begin{tabular}{|l|l|l|l|l|}
  \hline
    Model       & Test&Test&Dev&Dev \\
                & Clean&Other&Clean&Other\\
    \hline

    Hybrid Transformer+LM~\cite{wang2020b} &2.26&4.85&-&- \\
    RWTH+LM~\cite{rwth}&2.3&5.0&1.9&4.5\\
    \hline
    Baseline (L)+LM ~\cite{moritz2020} &2.7  &6.1   &2.4  &6.0\\
    ESPNet Transformer+LM ~\cite{karita2019asru}& 2.6  &5.7   & 2.2 &5.6\\
    LAS+LM~\cite{specaugment}&3.2&9.8&-&-\\
    LAS+SpecAugment+LM~\cite{specaugment}&2.5 &5.8 &- &- \\
    Transformer+LM~\cite{e2e2020}&2.33&5.17&2.10&4.79 \\
    Semantic Mask (L) ~\cite{wang2020a} &3.04 &7.43 & 2.93&7.75 \\ 
    \hspace{0.01cm} + LM, Speed P., Rescore~\cite{wang2020a}&2.24 &5.12 &2.05 &5.01 \\ \hline
    Baseline (Ours) &&&& \\
    \hline
    Transformer (S) &3.4 &8.3  &3.2  &8.5  \\
    \hspace{0.5cm}   + LM &2.3 &5.2 & 2.0&5.1\\
    Transfomrer (L) &3.0 &7.3  &2.8 &7.2  \\
    \hspace{0.5cm}   + LM &2.1 &5.0 &2.0 &4.8\\

\hline
    E2E ASR with GAN &&&& \\ \hline
    Transformer (S)+GAN&4.5 &11.1 &4.3 &11.4 \\
    \hspace{0.5cm}   + LM &2.6 & 5.6 &2.3 &5.4\\
    Transformer (S)+FTGAN&3.3 &7.9 &3.2 &8.2 \\
    \hspace{0.5cm}   + LM &2.2 &\textbf{4.7} &\textbf{1.9} &\textbf{4.7}\\
    Transformer (L) +FTGAN&\textbf{2.9} &\textbf{7.1} &\textbf{2.7} &\textbf{7.2} \\
    \hspace{0.5cm}   + LM &\textbf{2.1} &4.9 &\textbf{1.9} &\textbf{4.7}\\
\hline

    \end{tabular}
  \caption{WER results for E2E ASR models. (S) and (L) denote the small and the large settings respectively. +FTGAN represents the fine-tuning of pre-trained ASR model using GAN and +GAN describes the ASR model trained using GAN without pre-training ASR model.}
  \label{results}
\end{table}

\begin{table}[!htb]
  \small
  \centering
  \begin{tabular}{|l|l|l|l|l|}
  \hline
    1&2&3&4&5 \\

\hline
    0.9446827& 0.9444502& 0.9440999& 0.9434293&  0.9419135 \\ \hline
    0.9497483& 0.9489650&0.9485720& 0.9475869& 0.9473457 \\ \hline
     0.9277521& 0.9244638& 0.9241905&  0.9235519& 0.9225379 \\ \hline
    \textbf{0.9504781}& \textbf{0.9504015}& \textbf{0.9499697}& \textbf{0.9498766}& \textbf{0.9498552} \\ \hline
    \textbf{0.9535346}& \textbf{0.9535285}& \textbf{0.9535092}& \textbf{0.9531790} & \textbf{0.9531746} \\ \hline

    \end{tabular}
  \caption{Rows 2-6 represent the best five validation accuracy for the Transformer (S), Transformer (L), Transformer (S)+GAN, Transformer (S)+FTGAN,  and Transformer (L)+FTGAN  models respectively}
  \label{val5}
\end{table}
\vspace{-3mm}
\subsection{Results}
We show all of our experimental results in Table~\ref{results}. From Table~\ref{results},
we observe that our proposed fine-tuning approach of pre-trained ASR model using GAN (Transformer (S/L)+FTGAN) 
outperforms our baselines for both small (S) and large (L) models. Also, they show better results over RNN and other  transformer-based models~\cite{specaugment,karita2019asru,moritz2020, e2e2020,wang2020b}. Our large baseline model gives better test-clean and test-other WER results than the reported results in the ESPNET Github repository\footnote{ \url{https://github.com/espnet/espnet/blob/master/egs/librispeech/asr1/RESULTS.md}}. 
Also, our large baseline model outperforms a very recent work with semantic masking technique for transformer ASR~\cite{wang2020a}.
With LM fusion, our proposed fine-tuning approach gives the best results for both small and large settings. Our best WER results with LM fusion are \textbf{2.1\%}, \textbf{4.7\%}, \textbf{1.9\%} and \textbf{4.7\%} for the test-clean, test-other, dev-clean and dev-other respectively. Moreover, we perform an experiment using  small settings for adversarial training using GAN (Transformer (S)+GAN) without pre-training the ASR model to compare it with our fine-tuning approach. We train this model for 150 epochs to make equivalent number of epochs with our proposed approach (100 epochs for pre-training and 50 epochs for fine-tuning). Unlike~\cite{liu}, we see the performance drops over the baseline model. This might be because of adversarial training requires larger number of training epochs for large number of training samples~\cite{feizi}, high-variance gradient update from the discriminator and also the use of SpecAugmentation~\cite{specaugment}. On the other hand, our proposed fine-tuning approach using GAN can outperform the baseline model since we provide ASR output from a pre-trained ASR as an input to the discriminator which makes it harder for the discriminator to distinguish with the real data. Thus, the ASR model can further learn using the proposed fine-tuning approach by getting feedback from the discriminator through the adversarial loss.        

Moreover, we report the best five validation accuracy for the trained models in Table~\ref{val5}. We observe that our proposed fine-tuning approach using GAN (Transformer (S/L)+FTGAN) gives better validation accuracy for both small and large settings over the baseline models. Also, we note that adversarial training without pre-training ASR (Transformer (S)+GAN) yields worst validation accuracy  compared to the baseline model for the same reasons mentioned above.     

\vspace{-6pt}
\section{RELATION TO PRIOR WORK }
\label{sec:prior}
Attention-based encoder-decoder architectures for end-to-end  ASR models have  shown great success in the recent literature~\cite{chan2016listen,specaugment,karita2019asru}. Transformer architecture for end-to-end ASR has been explored extensively~\cite{karita2019asru,karita2019is} with employing convolution sub-sampling in the encoder pre-processing for efficient self-attention in the encoder. ~\cite{karita2019asru} focuses on multi-task learning with CTC, and show that transformer-based end-to-end ASR is highly competitive with RNN-based methods. In~\cite{wang2020a}, a semantic mask based regularization method was introduced for  transformer-based ASR to force the decoder to learn a better language model. A hybrid transformer model with deep layers  and iterated loss was introduced in~\cite{wang2020b}. In~\cite{e2e2020}, a semi-supervised learning with pseudo-labeling using transformer-based acoustic model was introduced.  In~\cite{semi}, a semi-supervised ASR model was developed using adversarial training with paired  low-resource corpus and non-parallel speech and text corpora. An unbalanced GAN was proposed for computer vision task in~\cite{vaegenerator}, where a variational auto-encoder (VAE) is trained first and then the weights of the decoder of the VAE is transferred as a pre-trained generator for the GAN training. In contrast, we don't use any external networks to pre-train our generator. Our ASR model is trained first and act as a generator for fine-tuning using adversarial approach. In~\cite{liu}, adversarial training for end-to-end ASR without pre-training was explored for small paired corpus with un-paired text data. 
In this work, we explored adversarial training of E2E ASR models using a large corpus without and with pre-training. We showed that, our proposed fine-tuning of pre-trained E2E ASR model using adversarial training approach outperforms the conventional adversarial training without pre-training~\cite{liu}. 

\vspace{-8pt}
\section{Conclusion and Future Work}
\label{sec:conclusion}
GAN-based adversarial training of end-to-end (E2E) ASR systems has explored recently for low-resource ASR corpora. Adversarial training of E2E ASR model using large ASR corpus with GAN framework may take excessively long time due to high-variance gradient updates. In this work, we proposed a fine-tuning approach of pre-trained ASR models with GAN-based adversarial training, where the ASR model acts as a generator and a discriminator tries to distinguish the ASR output transcriptions from the real data. As the ASR model is pre-trained, the discriminator cannot easily distinguish the ASR output from the real data which helps in training a  stronger discriminator, which in turn improves the performance of the ASR model during fine-tuning. The ASR model is trained with an additional adversarial loss against the discriminator and it learned so that it can fool the discriminator to reach the GAN equilibrium. Also, we run experiments for conventional adversarial training of E2E ASR in the GAN framework without pre-training an ASR model. 
Experiments on full LibriSpeech dataset showed that our proposed fine-tuning approach of pre-trained E2E ASR model outperforms the baseline and conventional adversarial training using GAN. For future work, we will apply VGG-like convolution sub-sampling~\cite{wang2020a} for further performance improvement. 


\bibliographystyle{IEEEbib}
\ninept
\bibliography{strings,refs}

\end{document}